\documentstyle[12pt]{article}
\newcommand{\be}{\begin{equation}}
\newcommand{\ee}{\end{equation}}
\newcommand{\nn}{\nonumber}
\newcommand{\beba}{\begin{equation}\begin{array}{lcl}}
\newcommand{\eaee}{\end{array}\end{equation}}
\newcommand{\bea}{\begin{eqnarray}}
\newcommand{\eea}{\end{eqnarray}}
\newcommand{\ba}{\begin{array}}
\newcommand{\ea}{\end{array}}

\newcommand{\ns}{\normalsize}
\newcommand{\refs}[1]{(\ref{#1})}
\def\a{\alpha}
\def\b{\beta}

\def\c{\chi}
\def\d{\delta}
\def\e{\epsilon}

\def\f{\phi}

\def\z{\psi}

\def\k{\kappa}

\def\m{\mu}
\def\n{\nu}

\def\p{\pi}
\def\q{\theta}

\def\r{\rho}
\def\s{\sigma}

\def\x{\xi}
\def\z{\zeta}

\def\D{\Delta}
\def\F{\Phi}
\def\G{\Gamma}
\def\J{\Psi}
\def\L{\Lambda}
\def\O{\Omega}

\def\ch{{\cal H}}
\def\cl{{\cal L}}

\def\tr{{\rm tr}}
\def\bz{{\bf 0}}
\def\bu{{\bf 1}}
\def\br{{\bf r}}
\begin{document}
\begin{titlepage}
\title{{\large\bf $O(d,d)$ Symmetry in Quantum Cosmology}\\
                          \vspace{-4cm}
                          \hfill{\ns TUM-HEP 237/96\\}
                          \hfill{\ns SFB-375/29\\}
                          \hfill{\ns hep-th/9602084\\[.5cm]}
                          \hfill{\ns January 1996}
                          \vspace{2cm} }

\author{Alexandros A. Kehagias\thanks{Supported by the Alexander von
                                      Humboldt--Stiftung}
        \thanks{Email : kehagias@physik.tu-muenchen.de}\\[0.2cm]
        {\ns and}\\[2mm]
        Andr\'e Lukas\thanks{Email : alukas@physik.tu-muenchen.de}
        \thanks{Address after March 1996~: Department of
        Physics, University of Pennsylvania,
        Philadelphia, PA 19104, USA}\\[1cm]
       {\ns Physik Department}\\
        {\ns Technische Universit\"at M\"unchen}\\
        {\ns D-85747 Garching, Germany}\\}

\date{}
\maketitle

\begin{abstract} \baselineskip=6mm
We analyze the quantum cosmology of one--loop string effective models
which exhibit an $O(d,d)$ symmetry. It is shown that due to the large
symmetry of these models the Wheeler--de Witt equation can completely
be solved. As a result, we find a basis of
solutions with well defined transformation properties under $O(d,d)$ and
under scale factor duality in particular. The general results are explicitly
applied to 2--dimensional target spaces while some aspects of
higher dimensional cases are also discussed. Moreover, a semiclassical
wave function for the 2-dimensional black hole is constructed as a
superposition of our basis.
\end{abstract}

\thispagestyle{empty}
\end{titlepage}

\section{Introduction}
String theory as a candidate for the fundamental theory of nature is
expected to predict not just the low energy effective particle theory but
the initial status of our universe~\cite{cosm}--\cite{mv2}. In any
case the prediction crucially depends on which of the perturbatively
degenerated vacua of string theory is chosen.
Nonperturbative effects should remove this degeneracy and pick
out the true vacuum. So far, however, the only information available
comes from nonperturbative effects like gaugino condensation~\cite{gau_cond}
which can be treated in the context of a low energy effective model. Therefore,
string cosmology at present cannot fully predict the initial state of the
universe. At most, the various promising candidate vacua can be checked
with respect to their cosmological acceptability.

Unfortunately, it has turned out that early universe cosmologies based on
gaugino condensate models have rather generic problems~\cite{br_stei,mod_prob}.
This has motivated the construction of alternative string cosmologies
e.~g.~based on the idea of topology change~\cite{topology} or the use of
stringy symmetries like duality~\cite{pbb}. The latter, so called
pre--big--bang cosmologies  relate the superinflationary
(pre--big--bang) and ordinary phase to each other by a duality
transformation~\cite{sfd}, also named scale factor duality.
The pre--big--bang models can be conveniently analyzed in the framework of
an $O(d,d)$--invariant one--loop string effective action~\cite{odd} with scale
factor duality as a discrete subgroup. For our purpose, namely to make
contact between string-- and quantum cosmology we will concentrate on this
interesting class of models.

Quantum Cosmology~\cite{hall} provides a completely different approach
towards a theory of initial conditions of the universe. Typically,
the Wheeler--de Witt equation is solved in minisuperspace i.~e.~for
a finite number of degrees of freedom. This is done, adopting a proposal for
the boundary conditions several of which have been advocated so far,
e.~g.~\cite{hh,vil}. A consistent interpretation of the resulting wave
function can be given in the semiclassical region and certain
predictions can be extracted in this limit.

Most probably, in the context of string theory such an approach does not
have any fundamental meaning though attempts have been made in this
direction~\cite{qc_st}.
Nevertheless one might ask the following questions~: What is the
``effective'' meaning of the quantum cosmological approach in string theory
and -- more pragmatically -- can the ``stringy'' properties of string
effective models be helpful to understand their quantum cosmology?
Certainly, the first of these questions will not be answered here though
our results might provide some hints for further study in this direction.
The answer to the second question will be definitely ``yes''.

Our study along these questions will be concentrated on the aforementioned
models with $O(d,d)$ symmetry for two reasons~: First of all, as discussed
before, cosmologically interesting models are included in this class of
string vacua so that relations between string-- and quantum cosmology
can be discussed. This has been done in ref.~\cite{bento_bert} for the
first time. Second, the large symmetry of these models turns out to be
very useful in solving their quantum cosmology. In particular, due to
the symmetry the system consists of a finite number of degrees of freedom
only so that we are dealing with a minisuperspace model. We will not attempt to
impose certain boundary conditions on our solutions following one of
the proposals in the literature. Instead, we will
give a complete classification of wave functions solving the
Wheeler--de Witt equation in terms of $O(d,d)$ quantum numbers and construct
some interesting semiclassical examples as appropriate superpositions.\\

The plan of the paper is as follows: In section 2 we present a short review
of the $O(d,d)$--invariant effective action, its properties and the
Hamiltonian formulation. Section 3 deals with the
operator realization of $O(d,d)$ which is used to calculate the quadratic
$O(d,d)$ Casimir. These results are applied in section 4 to find the
general solution of the Wheeler--de Witt equation for arbitrary dimension.
The 2--dimensional case is discussed in section 5 where some
semiclassical wave functions are also constructed using our basis of solutions.
Furthermore, we consider some aspects of higher dimensional cases,
in particular the 3-- and 4--dimensional ones. Relevant $O(d,d)$-group
theory has been summarized in the appendix.
\section{Setup of general formalism}
The bosonic part of the one--loop string effective action~\cite{eff_action}
reads
\be
 S = \int\; d^Dx\;\sqrt{-g}\, e^{-\f}\left[ \L -R-g^{\m\n}
      \partial_\m\f\partial_\n\f
     -\frac{1}{12}h_{\m\n\r}h^{\m\n\r}\right]\; , \label{action}
\ee
where $\L$ is the cosmological constant, $\f$ the dilaton,
$g_{\m\n}$ the $\s$--model metric and the torsion $h_{\m\n\r}$ is given
in terms of the antisymmetric field $b_{\m\n}$ as
\be
 h_{\m\n\r} = \partial_\m b_{\n\r}+\partial_\n b_{\r\m}+
              \partial_\r b_{\m\n}\; .
\ee
If $g$ and $b$ do not depend on $d$ of the $D$ space--time coordinates the
above action turns out to have an $O(d,d)$ symmetry~\cite{nar}. For
cosmological considerations we assume that all fields in eq.~\refs{action}
are time dependent only ($d=D-1$). By a suitable coordinate and gauge
transformation $g$ and $b$ can be brought to the block diagonal form
\be
 g = \left(\ba{cc} -N^2(t)&\bz\\\bz&G(t)\ea\right)\; ,\quad\quad
 b = \left(\ba{cc} \bz&\bz\\\bz&B(t)\ea\right)\; , \label{gb}
\ee
where $G$, $B$ are $d\times d$ matrices and we have also introduced a lapse
function $N$. Using the above form of the metric and the antisymmetric
tensor, the action~\refs{action} turns out to be
\bea
 S &=&\int dt\; N\sqrt{{\rm det}(G)}\; e^{-\f}\left[\L +\frac{1}{N^2}\left(
   \dot{\f}^2-\dot{\f}\;\tr (G^{-1}\dot{G})+\right.\right.\nn \\
   &&\quad\quad\left.\left.\frac{1}{4}(\tr (G^{-1}\dot{G}))^2
   \right)+\frac{1}{4N^2}\tr\left( (G^{-1}\dot{B})^2-(G^{-1}\dot{G})^2\right)
   \right]\; . \label{oddact}
\eea
The $O(d,d)$ invariance of eq.~\refs{oddact} can be made explicit by
introducing the matrix
\be
 M=\left(\ba{cc}G^{-1}&-G^{-1}B\\BG^{-1}&G-BG^{-1}B\ea\right)
\ee
which transforms as
\be
 M\rightarrow \O^T M\O
\ee
under the action of an $O(d,d)$--element $\O$ which, by definition, satisfies
\be
 \O^T\eta\O = \eta\; ,\quad\quad\eta =\left(\ba{cc}\bz&\bu\\\bu&\bz
              \ea\right)\; . \label{odd_def}
\ee
In particular, the $O(d,d)$--element $\O = \eta$ generates
$T$--duality transformations which are direct generalizations of the
$R\rightarrow 1/R$ symmetry of the toroidally compactified string.
Moreover, the matrix $M$ is an $O(d,d)$ element itself which is
restricted to be symmetric, i.~e.,
\be
 M^T\eta M = \eta\; ,\quad\quad M=M^T\; . \label{constraints}
\ee
We may express the action~\refs{oddact} in a manifest $O(d,d)$--invariant
form as
\be
 S=\int\; dt\; Ne^{-\F}\left[\L +\frac{1}{N^2}\dot{\F}^2+\frac{1}{8N^2}
   \tr\left (\dot{M}^2\eta )^2\right)\right]\; ,
 \label{man_act}
\ee
by using the $O(d,d)$--invariant dilaton
\be
 \F = \f -\ln\sqrt{{\rm det}(G)}
\ee
and the relation
\be
 \tr\left( (\dot{M}\eta )^2\right) = 2\,\tr\left[\left( G^{-1}\dot{B}\right)^2
    -\left( G^{-1}\dot{G}\right)^2\right]\; .
 \label{mgb}
\ee
The classical solutions of the action~\refs{man_act},  classified
in terms of a ``$\b$--function'' which determines the running of $e^\F$,
may completely be expressed in terms of quadratures~\cite{mv2}.

Turning to the quantum theory now we may employ the Hamiltonian formulation.
The action~\refs{man_act} defines a
constrained system with the constraints given by eq.~\refs{constraints}.
Its quantization would therefore lead to unnecessary complications which
can be avoided by using the fields $G$, $B$ instead of the
$O(d,d)$--covariant object $M$. Nevertheless, the invariant combination
$\F$ turns out to be useful in the following. Consequently, we have chosen,
as the starting point, the field basis $\{ \F ,G,B \}$ and the
corresponding Lagrangian
\be
 \cl = Ne^{-\F}\left[\L+\frac{1}{N^2}\dot{\F}^2+\frac{1}{4N^2}\tr\left(
       (G^{-1}\dot{B})^2-(G^{-1}\dot{G})^2\right)\right]\; .
\ee
The conjugate momenta are
\bea
 \p_\F &=& \frac{2}{N}e^{-\F}\dot{\f}\nn\\
 \p_G &=& -\frac{1}{N}e^{-\F}G^{-1}\dot{G}G^{-1}\label{momenta}\\
 \p_B &=& -\frac{1}{N}e^{-\F}G^{-1}\dot{B}G^{-1}\nn\; ,
\eea
the Hamiltonian turns out to be
\be
 \ch = N\left[\frac{e^\F}{4}\left(\p_\F^2 +\tr\left( (G\p_B)^2-(G\p_G)^2\right)
       \right) -e^{-\F}\L\right]\; ,
 \label{ham}
\ee
and the canonical quantization relations are
\bea
 \left[\F ,\p_\F\right] &=& i\nn \\
 \left[ G_{ij},\p_{G_{kl}}\right] &=& i(\d_{ik}\d_{jl}+\d_{il}\d_{jk})\\
 \left[ B_{ij},\p_{B_{kl}}\right] &=& i(\d_{ik}\d_{jl}-\d_{il}\d_{jk})\nn \; .
\eea
The wave function $\J = \J (\F ,G,B)$ is subject to the Wheeler--de Witt
equation
\be
 \hat{\ch}\J = 0 \label{wdw}
\ee
with the operator version $\hat{\ch}$ of the Hamiltonian~\refs{ham}.
The above equation is a zero energy Schr\"odinger equation which -- due to
the restriction of $O(d,d)$ invariance -- contains a finite number
of degrees of freedom only. In the language of Quantum Cosmology it
constitutes a minisuperspace model~\cite{hall}. Turning the
Hamiltonian~\refs{ham} into $\hat{\ch}$ we encounter the familiar
operator--ordering ambiguity which appears in the term
$\tr (G\p_G )^2$~\cite{opord}. Later on we will comment on this ambiguity
and we will show how to partially fix it.

\section{Operator realization of the $O(d,d)$ symmetry}
Classically, to any $O(d,d)$ generator $T$ corresponds a conserved current
which turns out to be
\be
 J_T=e^{-\F}\tr\left( T\eta\dot{M}\eta M\right)\; .
 \label{J_def}
\ee
Working in the standard basis~\refs{odd_basis} of matrices $(T_{AB})$, these
currents take the form~\footnote{Group properties of $O(d,d)$ which are
relevant for the following discussion are collected in the appendix.}
\be
 \left( J_{AB}\right) = \left( \ba{cc} S&C\\D&R \ea\right) \label{jab}
\ee
with
\bea
 S &=& \p_B \nn \\
 C &=& \p_G G -\p_B B \nn\\
 D &=& -(G\p_G -B\p_B) \label{class_curr}\\
 R &=& G\p_G B - B\p_B B+B\p_G G -G\p_B G \nn \; .
\eea
It can be easily seen that $S$ generates the shifts of the antisymmetric
tensor field whereas $C$ generates coordinate transformations. They
correspond to the explicit transformations given in eq.~\refs{special_trafo}.

To promote these currents to their operator version we have to
carefully handle the operator--ordering problem. Let us define the operators
\bea
 \hat{S} &=& S \nn \\
 \hat{C} &=& C+ic\bu \\
 \hat{R} &=& R+irB \nn
\eea
with an operator ordering in $S,C,R$ defined by the order
of matrices as in eq.~\refs{class_curr}, while the real quantities $c,r$
parameterize the operator--ordering ambiguity. In the standard basis
we can write
\be
 \left( \hat{J}_{AB}\right) = \left( \ba{cc} \hat{S}&\hat{C}\\
                              -\hat{C}^T&\hat{R} \ea\right)\; .
\ee
By an explicit calculation we find that these operators satisfy the
$O(d,d)$ algebra~\footnote{Lower case letters $i,j,k,\cdots = 1,\cdots ,d$
refer to one of the $O(d)$ subgroups or a spatial coordinate whereas upper
case letters $A,B,C,\cdots = 1,\cdots , 2d$ denote $O(d,d)$ group indices.}
\bea
 \left[ \hat{S}_{ij},\hat{S}_{kl}\right] &=& 0\nn\\
 \left[ \hat{S}_{ij},\hat{C}_{kl}\right] &=& i(\d_{il}\hat{S}_{jk} -
\d_{jl}\hat{S}_{ik})\nn\\
 \left[ \hat{S}_{ij},\hat{R}_{kl}\right] &=&
i(\d_{jk}\hat{C}_{il}+\d_{il}\hat{C}_{jk}
                                 -\d_{ik}\hat{C}_{jl}-\d_{jl}\hat{C}_{ik})
 \label{algebra} \\
 \left[ \hat{C}_{ij},\hat{C}_{kl}\right] &=&
i(\d_{jk}\hat{C}_{il}-\d_{il}\hat{C}_{kj})\nn\\
 \left[ \hat{C}_{ij},\hat{R}_{kl}\right] &=&
i(\d_{il}\hat{R}_{jk}-\d_{ik}\hat{R}_{jl})\nn \\
 \left[ \hat{R}_{ij},\hat{R}_{kl}\right] &=& 0\nn \; ,
\eea
if and only if
\be
 2c-r = 2d\; . \label{ord_cond}
\ee
If the condition~\refs{ord_cond} is violated, however, unwanted central
terms appear. Next we want to calculate the quadratic
Casimir~\footnote{If not specified explicitly, we refer to operators from
now on and we drop the hat notation.}
\be
 C = -\frac{1}{16}f_{(AB)(EF)(GH)}f_{(CD)(GH)(EF)}J_{AB}J_{CD}
\ee
in terms of the operator representation $J_{AB}$. With the structure
constants to be read off from eq.~\refs{odd_alg} we arrive at
\be
 C = -2(d-1)\tr\left[ (J\eta )^2\right] \; .
 \label{casimir}
\ee
A straightforward calculation using the explicit expressions for $J$ and
eq.~\refs{ord_cond} leads to
\bea
 C &=& 4(d-1)\left[\tr\left( (G\p_B )^2-(G\p_G )^2\right)\right.\nn \\
    && \left. +i(3d-2c+1) \tr\left(\p_G G\right) +dc(c-d+2)\right]\; .
       \label{op_casimir}
\eea
The parameter $c$ is still arbitrary and can e.~g.~be chosen such that one
of the two extra terms in $C$ disappears. We observe that the first term
exactly coincides with the $G,B$-dependent part of the classical
Hamiltonian~\refs{ham}. To make $\ch$ and $C$ fully compatible
an appropriate operator ordering prescription for the Hamiltonian can
be chosen and in the following we assume that this has been done. Then the
ambiguity in promoting the classical Hamiltonian to its quantum version is
partially fixed. We stress that such a choice does not exist if
the condition~\refs{ord_cond} was violated or, equivalently, $O(d,d)$
symmetry was broken. In that case the expression
for $C$ contains a term proportional to $\tr (\p_B B)$ which never can
appear in the Hamiltonian.

\section{Solving the Wheeler--de Witt equation}
Now we are ready to solve the Wheeler--de Witt equation exploiting
the $O(d,d)$ symmetry. According to the above remarks, the $G,B$-dependent
part of the Hamiltonian can be chosen to coincide with the Casimir so that
\be
 \ch = N\left[ \frac{e^\F}{4}\left(\p_\F^2 +\frac{1}{4(d-1)}C\right)
       -e^{-\F}\L\right]\; ,\quad d>1\; .
\ee
We emphasize that $\ch$ can be entirely expressed in terms of the quadratic
Casimir. No higher Casimir operators or ``radial'' parts are present.
Obviously, the above expression does not hold for the abelian case $d=1$
since the Casimir vanishes. It has to be replaced by the $O(1,1)$--charge
$Q=iG\p_G$ leading to the Hamiltonian
\be
 \ch = N\left[ \frac{e^\F}{4}\left(\p_\F^2 +Q^2\right)
       -e^{-\F}\L\right]\; ,\quad d=1\; .
 \label{d1_ham}
\ee
Note that the definition of $Q$ fixes the operator ordering for $d=1$.

Via the ansatz $\J (\F ,G,B) =\x (\F )\c (G,B)$, the Wheeler--de Witt equation
can be separated into a $G,B$--dependent part
\be
 E\c = \left\{\ba{ccc} \frac{C}{4(d-1)}\c&,&d>1\\
                     Q^2\c&,&d=1\ea\right.
 \label{dgl_gb}
\ee
and a dilaton part
\be
 \left(\p_\F^2 -4e^{-2\F}\L+E\right)\x = 0\; .\label{deq_phi}
\ee
The first equation~\refs{dgl_gb} specifies the spectral parameter $E$ in terms
of the Casimir $C$. We will concentrate on the finite
dimensional representations of $O(d,d)$ which can be directly obtained
from the corresponding $O(2d)$ representations by multiplying the
noncompact generators with $i$ (Weyl's trick). Referring to these
representations by Dynkin labels~\cite{slansky}
$(\br )= (a_1,\cdots ,a_d)$, the Casimir is
\be
 C(\br ) = (\br ,\br +\d)\; ,\quad\quad \d = (2,\cdots ,2)
 \label{cas_val}
\ee
where the product $(..,..)$ is calculated with the metric tensor of $SO(2d)$.
The solutions $\c$ are harmonic functions on the moduli space
$O(d,d)/O(d)\times O(d)$. To find these harmonics we have to concentrate on
those representations $(\br )$ of $O(2d)$ which in their decomposition
under $O(d)\times O(d)$ contain a singlet. Let us denote weights in a certain
representation $(\br )$ by $(m) =(m_1,\cdots ,m_d)$
and the representation matrices  by $D^{(\br )}_{(m)(m')}$.
Then the $G,B$--dependent part of the wave function takes the form
\be
 \c (G,B) \sim D^{(\br )}_{(0\z )(m)}\left( M(G,B)\right)
 \label{gb_sol}
\ee
where $(0\z )$ denotes the $O(d)\times O(d)$ singlet directions.
For representations which
can be obtained as tensor products of the fundamental representation of
$O(d,d)$ there is a simple way to get explicit expressions for $\c$.
They can be written as the following products of matrix elements $M_{AB}$~:
\be
 \c\sim (M_+M_+^T)_{A_1B_1}\cdots (M_+M_+^T)_{A_mB_m}
        (M_-M_-^T)_{C_1D_1}\cdots (M_-M_-^T)_{C_nD_n}\; .
\ee
Here $M_\pm$ denote the matrices consisting of the $d$ upper or lower
rows of $M$, respectively. The indices $A_k,B_k,C_k,D_k$ have to be
symmetrizes appropriately in accordance with the desired
$O(d,d)$ representation.

The equation~\refs{deq_phi} for $\F$ becomes simple for a vanishing
cosmological constant. For $\L$ nonvanishing and by using the substitution
\be
 \r = 2\sqrt{|\L |}e^{-\F}\; ,
\ee
it can be rewritten as a Bessel differential equation.
The solution may then be expressed as
\be
 \x = \left\{\ba{ccc}
      ae^{\sqrt{E}\F}+be^{-\sqrt{E}\F}&,&\L =0\\
      aJ_{\sqrt{E}}(\r )+bN_{\sqrt{E}}(\r )&,&\L >0\\
      aJ_{\sqrt{E}}(i\r )+bN_{\sqrt{E}}(i\r )&,&\L <0\ea\right.\; ,
     \label{phi_sol}
\ee
in terms of the Bessel-- and von Neumann--functions $J_\n$, $N_\n$ and
arbitrary coefficients $a,b$.
The expressions~\refs{gb_sol} and~\refs{phi_sol} with the
Casimir~\refs{cas_val}, taken for all $O(2d)$ representations
which contain $O(d)\times O(d)$ singlets, form a complete set of solutions
of the Wheeler--de Witt equation. Explicit decompositions of certain
semiclassical wave functions in terms of this basis will be given in the
next section.
\section{Application to 2--dimensional models}
In this section we apply our results to 2--dimensional models
in order to find explicit expressions for our solutions.
Furthermore, we show how to construct semiclassical wave functions
by appropriate superpositions and discuss some properties of the
higher dimensional cases.

Clearly, for $d=1$ the situation simplifies a lot. The matrix $M$ is of the
form
\be
 M = \left(\ba{cc} G^{-1}&0\\0&G\ea\right)\; ,
\ee
where $G$ is a single degree of freedom from the metric and obviously
no antisymmetric tensor field exists. Of course the
Wheeler--de Witt equation can be integrated immediately in that case.
To illustrate the methods presented above, however, we follow the group
theoretical approach here.

The relevant group $O(1,1)$ can be parameterized as
\be
 \O (\a ) = \left(\ba{cc} \cosh\frac{\a}{2}&\sinh\frac{\a}{2}\\
                          \sinh\frac{\a}{2}&\cosh\frac{\a}{2}\ea\right)
 \label{o11}
\ee
and its representations can be labeled by a charge $q$~:
\be
 D^{(q)}(\O (\a )) = e^{q\a}\; .
\ee
The above parameterization refers to the basis where
$\eta$ is diagonal. In this basis the matrix $M$ takes the form
\be
 M = \frac{1}{2}\left(\ba{cc} G+G^{-1}&G-G^{-1}\\G-G^{-1}&G+G^{-1}\ea\right)
\ee
and by comparison with eq.~\refs{o11} one reads off that $G=e^{\a /2}$.
Therefore $\c\sim G^{q/2}$ and the complete wave function can be written as
\be
 \J^{(q)}(G,\F ) = G^{q/2}\left\{\ba{lcc}
                   ae^{|q|\F}+be^{-|q|\F}&,&\L = 0\\
                   aJ_{|q|}\left( 2\L e^{-\F}\right) +
                   bN_{|q|}\left( 2\L e^{-\F}\right)&,&\L > 0\\
                   aJ_{|q|}\left( i2\L e^{-\F}\right) +
                   bN_{|q|}\left( i2\L e^{-\F}\right)&,&\L < 0\ea\right.\; .
 \label{d1_sol}
\ee
As we will  see below,  duality acts on the representations by complex
conjugation, i.~e.~by $q\rightarrow -q$ in the case under consideration.
Indeed, eq.~\refs{d1_sol} shows that this transformation is equivalent
to $G\rightarrow G^{-1}$.\\

Next, we construct semiclassical wave functions as superpositions
of the solutions~\refs{d1_sol}, i.~e.~wave functions of the form
\be
 \J = C e^{iS}
\ee
with a classical action $S$ and a (real) prefactor $C$. Following the
interpretation that a peak in the wave function is considered as a
prediction~\cite{peak} it can be shown~\cite{hall2} that such a semiclassical
wave function selects a subset of classical trajectories specified by
\be
 \p_G = \frac{\partial S}{\partial G}\; ,\quad\quad
 \p_\F = \frac{\partial S}{\partial\F}\; .
 \label{subset}
\ee
Here $\p_G$ and $\p_\F$ denote the classical expressions for the conjugate
momenta. On this subset the square of the wave function $|\J |^2=C^2$ can be
consistently interpreted as a probability~\cite{hall}.

Let us start with the case of vanishing cosmological constant $\L =0$.
For a given charge $q$ we have the solution $(G^{\pm 1/2}e^{-\F})^{|q|}$
where the sign is specified by the sign of $q$. This implies that we can
easily construct a semiclassical wave function with
\be
 S_\pm \sim G^{\pm 1/2}e^{-\F}
\ee
which by inspections of eqs.~\refs{subset} corresponds to a Milne universe
and its dual~\cite{myer,milne}, i.~e.~$G\sim t^{\mp 2}$ and $\F\sim\ln t$.

For positive cosmological constant an approximate wave function in
the weak coupling region $e^{-\F}\gg 1$ can be obtained from the asymptotic
expansion of the Bessel function. For a fixed charge $q$ we find
\be
 S\simeq 2\sqrt{\L}\, e^{-\F}\; ,\quad\quad
 C\simeq\sqrt{\frac{1}{\p\sqrt{\L}}}\, G^{q/2}e^{\F /2}\; .
\ee
This implies a set of classical solutions specified by $\dot{\F} =-\sqrt{\L}$
and $\dot{G}=0$ which describe a static universe with a linearly moving
dilaton~\cite{myer}. It also corresponds to the linear dilaton
solution~\cite{lin_dil} in 2--d black hole physics. Depending on the sign
of $q$ the probability $C^2$ shows a preference for large or small static
universes and both cases are related to each other by
$R\rightarrow 1/R$ duality.

In addition to the linear dilaton solution we should also be able to find
a semiclassical black hole wave function. The corresponding classical
solution reads~\cite{bh,mv2}
\be
 G=\tanh^{\pm 2}\left(\sqrt{\L}(T-t)/2\right)\; ,\quad\quad
 e^\F = \frac{2\sqrt{\L}}{\k\sinh\left(\sqrt{\L}(T-t)\right)}\; ,
\ee
with arbitrary constants $\k ,T$. Its classical action can be computed from
\be
 S = 2\L\int dt\; e^{-\F}
\ee
and leads to the semiclassical wave function
\be
 \J = \exp\left[ i\sqrt{\L}\, e^{-\F}\left(\sqrt{G}+\frac{1}{\sqrt{G}}
      \right)\right]\; ,
 \label{bh_wf}
\ee
which  shows invariance under duality. Let us concentrate on the
particular solution $\J_+^{(q)}=G^{q/2}J_{|q|}(2\sqrt{\L}e^{-\F})$ taken
from eq.~\refs{d1_sol}.
We can use the Laurent series which generates Bessel functions~\cite{abr}
to arrive at the result
\be
 \J = \J_+^{(0)}+\sum_{q=1}^{\infty}i^q\left(\J_+^{(q)}+\J_+^{(-q)}\right)\; .
\ee
The absence of any prefactor in addition to eq.~\refs{bh_wf} shows that $\J$
-- though computed semiclassically -- is an exact solution of the
Wheeler--de Witt equation. Of course this can also be confirmed by direct
calculation if the operator ordering fixed by eq.~\refs{d1_sol} is used.\\

An advantage of our choice of basis is clearly its well defined
transformation property under $O(d,d)$. Though explicit expressions are more
difficult to find in higher dimensions this allows to discuss some
properties of our solution using group theoretical methods.
For $d=2$, e.~g.~we have to consider representations of
$SO(4)\simeq SU_L(2)\times SU_R(2)$ which are specified by
$(\br )=(a_L,a_R)=2(j_L,j_R)$. The Casimir is given by
\be
 C_{SO(4)}((j_L,j_R)) = 2(j_L(j_L+1)+j_R(j_R+1))
\ee
Locally, the moduli space can be written as
$SU_L(2)\times SU_R(2)/U_L(1)\times U_R(1)$
and the representations which contain a $U_L(1)\times U_R(1)$ singlet are
just the vector representations $(j_L,j_R)$ with $j_L,j_R$ integer.

For $d=3$ the representations of $SO(6)\simeq SU(4)$ are labeled by
$(\br )=(a_1,a_2,a_3)$. With the metric tensor
\be
 G_m(SU(4)) = \frac{1}{4}\left(\ba{ccc}3&2&1\\2&4&2\\1&2&3\ea\right)
\ee
we get for the Casimir
\bea
 C_{SU(4)}((a_1,a_2,a_3)) &=& \frac{3}{4}a_1^2+3a_1+a_1a_2+4a_2+\frac{1}{2}
                            a_1a_3\nn \\
                          &&+3a_3+a_2^2+a_2a_3+\frac{3}{4}a_3^2\; .
\eea
For a given representation $(\br )=(a_1,a_2,a_3)$ the complex conjugate
representation is specified by the label $(\br^* )=(a_3,a_2,a_1)$.
These are exactly the representations which are mapped into each other by
duality. From the local form of the moduli space $SU(4)/SU(2)\times SU(2)$
and the weight projection matrix
\be
 P(SU(4)\rightarrow SU(2)\times SU(2)) = \left(\ba{ccc} 1&1&0\\0&1&1\ea
                                         \right)\; ,
\ee
it can be seen that the representations containing
$SU(2)\times SU(2)$ singlets are just the one with $a_1+a_3$ even.

Finally, let us denote how duality acts on our solutions.
One may see from eq.~\refs{car_trans} that a state of weight $(m)$ in a
given representation $(\br )$ of $O(d,d)$ is mapped to a state of
weight $(-m)$. Correspondingly, the representation $(\br )$ is mapped into
its complex conjugate $(\br^*)$. This implies for our wave functions that
\bea
 \c (G,B) \sim D^{(\br )}_{(m)(0)}\left( M(G,B)\right)&\rightarrow&
          D^{(\br^* )}_{(-m)(0)}\left( M(G,B)\right)\nn \\
 \x (\F)&\rightarrow&\x (\F) \; .
\eea
\section{Conclusions}

Above, we have discussed quantum cosmological aspects of the one-loop string
effective action. Although, string theory is supposed to be a consistent theory
for quantum gravity, we believe that
it is meaningful to discuss quantum cosmology
in the context of the effective theory. Especially, as long as the
information about nonperturbative string theory is very limited, such an
effective approach may provide us with some hints about the correct
vacuum of the theory.

In our case, we have considered
cosmological backgrounds where all fields, namely, the metric, antisymmetric
tensor and the dilaton are present. In particular, we have assumed homogeneity
but not isotropy of the D-dimensional space-time. Moreover, the
surfaces of simultaneity are assumed to be flat so that we are actually
discussing here the quantum cosmology of the rolling--moduli
solution~\cite{myer}. The central point is the existence of an $O(d,d)$
symmetry of the one-loop beta-function equations. This symmetry allowed us
to separate and finally solve the Wheeler--de Witt equation in minisuperspace.
Although the general solution is quit complicated, it is possible to discuss
explicitly the lower-dimensional (D=2) case. Here, we have found all the known
2-dimensional solutions, namely, the Milne universe, the linear dilaton
solution as well as the 2-dimensional black hole one, depending on the value of
the cosmological constant. It should be noted that these solutions are the only
ones which have a CFT description, at least for specific (discrete) values of
$\Lambda$.

Since we are dealing with noncompact space--times we have not attempted to
impose one of the boundary proposals of quantum cosmology on our solutions.
Therefore, we are not able to specify e.~g.~the ground state wave function
selected by these proposals. However, one might speculate about another way to
choose a ground state, namely by demanding that it is invariant
under (scale factor) duality. In the 2--dimensional case this implies
a vanishing $O(1,1)$ charge $q$ and an essentially unique wave function
which depends on the invariant dilaton $\F$ only.

As a final comment let us note that one may consider higher--loop corrections
to the one-loop effective action as perturbations in the Wheeler--de Witt
equation  and use standard perturbation theory to deal with them.
\\

{\bf Acknowledgment} A.~L. would like to thank J.~E. Kim, K. Choi and
in particular E.~J. Chun for hospitality during his stay in Korea where
part of this work was done. This work was supported by the EC under
contract no.~SC1-CT92-0789 and by the Sonderforschungsbereich 375--95
``Research in Astroparticlephysics'' of DFG.
\renewcommand{\theequation}{\thesection.\arabic{equation}}
\newcommand{\sect}[1]{\section{#1}\setcounter{equation}{0}}
\section*{Appendix}
\appendix
\sect{Some useful properties of $O(d,d)$}
Our final goal is to find the solutions of the Wheeler--de Witt
equation~\refs{wdw} classified according to $O(d,d)$ quantum numbers.
As a preparation, we will now present a brief summary of the
$O(d,d)$ algebra and discuss its operator realization~\cite{giveon}.

By means of eq.~\refs{odd_def} a generator $T$ corresponding to the
infinitesimal transformation $\O = \bu +iT$ is subject to the constraint
\be
 T^T\eta +\eta T = 0\; .
\ee
Therefore the Lie algebra is given by matrices $T$ of the form
\be
 T = \left(\ba{cc} a&b\\c&-a^T \ea\right)
\ee
with $d\times d$ matrices $a$, $b$, $c$. Here $a$ is arbitrary and
$b$, $c$ are antisymmetric. The above form implies that $T\eta$ is
a generally antisymmetric matrix.
Infinitesimal shifts of the $B$-field $T_\q$ and coordinate transformations
$T_a$ are generated by
\be
 T_\q = \left(\ba{cc} \bz&\q\\\bz&\bz\ea\right)\; ,\quad\quad
 T_a = \left(\ba{cc} -a^T&\bz\\\bz&a\ea\right)\; , \label{special_trafo}
\ee
respectively.
As a standard basis for the Lie algebra we choose the matrices
\be
 \left( T_{AB}\right)^C_D = i\left( \d^C_A\eta_{BD}-\d^C_B\eta_{AD}\right)
 \label{odd_basis}
\ee
which satisfy the well--known commutation relations
\be
 \left[ T_{AB},T_{CD}\right] = i\left(-\eta_{AC}T_{BD}-\eta_{BD}T_{AC}
                               +\eta_{AD}T_{BC}+\eta_{BC}T_{AD}\right)\; .
 \label{odd_alg}
\ee
Let us also explicitly perform the Cartan decomposition of $O(d,d)$.
Here we switch to a different basis with
\be
 \eta = \left(\ba{cc} \bu&\bz\\\bz&-\bu \ea\right)\;
\ee
Defining the elements of the Cartan subalgebra as
\be
 H_i = T_{i+d,i}
 \label{cart_sub}
\ee
and raising and lowering operators as
\bea
 E_{ij}^{++} &=& \frac{1}{2}(T_{ij}+T_{i+d,j+d}+T_{i,j+d}+T_{i+d,j})\nn \\
 E_{ij}^{--} &=& -\frac{1}{2}(T_{ij}+T_{i+d,j+d}-T_{i,j+d}-T_{i+d,j})
 \label{cart_decomp} \\
 E_{ij}^{+-} &=& \frac{1}{2}(T_{ij}-T_{i+d,j+d}-T_{i,j+d}+T_{i+d,j})\nn \\
 E_{ij}^{--} &=& -\frac{1}{2}(T_{ij}-T_{i+d,j+d}+T_{i,j+d}-T_{i+d,j})\nn
\eea
we indeed find
\bea
 \left[ H_i,H_j\right] &=& 0 \nn \\
 \left[ H_i,E_{kl}^{(\e_1\e_2)}\right] &=& \a^{(\e_1\e_2,kl)}_i
                                           E_{kl}^{\e_1\e_2)} \\
 \left[ E_{ij}^{(++)},E_{ij}^{(--)}\right] &=& \sum_k \a^{(++,ij)}_kH_k\nn \\
 \left[ E_{ij}^{(+-)},E_{ij}^{(-+)} \right] &=& \sum_k \a^{(+-,ij)}_kH_k\nn
\eea
with the $SO(2d)$ roots
\be
 \a^{(\e_1\e_2,kl)}_i = \e_1\d^k_i+\e_2\d^l_i
\ee
and $\e_1,\e_2 =\pm$.
It is interesting to work out how duality acts on these generators.
{}From the transformation property $T_{AB}\rightarrow \eta T_{AB}\eta$ of
$T_{AB}$ and the definitions~\refs{cart_sub} and~\refs{cart_decomp} we find
\bea
 H_i &\rightarrow& -H_i \nn \\
 E_{ij}^{(++)}&\rightarrow& -E_{ij}^{(--)} \label{car_trans}\\
 E_{ij}^{(+-)}&\rightarrow& -E_{ij}^{(-+)}\nn\; .
\eea
\end{document}